\documentclass[fleqn,10pt]{wlscirep}
\usepackage[T1]{fontenc}

\usepackage[%
  autocite    = superscript,
  backend     = biber,
  sortcites   = true,
  style       = nature,
]{biblatex}
\bibliography{bibfile}
\newcommand{\authorcite}[1]{\citeauthor{#1}\,\supercite{#1}}

\usepackage{authblk}
\usepackage{graphicx}
\usepackage{amsmath}
\usepackage{placeins}
\usepackage[doublespacing]{setspace}
\usepackage{todonotes}
\usepackage{xr-hyper}
\usepackage[normalem]{ulem} %
\title{Network constraints on learnability of probabilistic motor sequences}

\author[1,2,3]{Ari E. Kahn}
\author[4]{Elisabeth A. Karuza}
\author[3,5,2]{Jean M. Vettel}
\author[2,6,7,8*]{Danielle S. Bassett}
\affil[1]{Department of Neuroscience, University of Pennsylvania, Philadelphia, PA 19104 USA}
\affil[2]{Department of Bioengineering, University of Pennsylvania, Philadelphia, PA 19104 USA}
\affil[3]{Human Research and Engineering Directorate, U.S. Army Research Laboratory, Aberdeen, MD 21001 USA}
\affil[4]{Department of Psychology, University of Pennsylvania, Philadelphia, PA 19104 USA}
\affil[5]{Department of Psychological and Brain Sciences, University of California, Santa Barbara, CA 93106 USA}
\affil[6]{Department of Electrical \& Systems Engineering, University of Pennsylvania, Philadelphia, PA 19104 USA}
\affil[7]{Department of Neurology, Perelman School of Medicine, University of Pennsylvania, Philadelphia, PA 19104 USA}
\affil[8]{Department of Physics \& Astronomy, College of Arts and Sciences, University of Pennsylvania, Philadelphia, PA 19104 USA}
\affil[*]{To whom correspondence should be addressed: dsb@seas.upenn.edu.}

\begin{document}

\maketitle

\section*{Keywords:} graph learning, statistical learning, motor sequence learning, graph theory, probabilistic sequences

\section*{Abstract}
Human learners are adept at grasping the complex relationships underlying incoming sequential input \autocite{Aslin2012}. In the present work, we formalize complex relationships as graph structures \autocite{newman2010networks} derived from temporal associations \autocite{Schapiro2013,karuza2017process} in motor sequences. Next, we explore the extent to which learners are sensitive to key variations in the topological properties \autocite{newman2011complex} inherent to those graph structures. Participants performed a probabilistic motor sequence task in which the order of button presses was determined by the traversal of graphs with modular, lattice-like, or random organization. Graph nodes each represented a unique button press and edges represented a transition between button presses. Results indicate that learning, indexed here by participants' response times, was strongly mediated by the graph's meso-scale organization, with modular graphs being associated with shorter response times than random and lattice graphs. Moreover, variations in a node's number of connections (degree) and a node's role in mediating long-distance communication (betweenness centrality) impacted graph learning, even after accounting for level of practice on that node. These results demonstrate that the graph architecture underlying temporal sequences of stimuli fundamentally constrains learning, and moreover that tools from network science provide a valuable framework for assessing how learners encode complex, temporally structured information.
\section*{Main}

Our ability to interact with our environment necessitates that we parse complex stimuli into smaller units, such as words and phrases in language input, or events in streams of visual stimuli. This essential process relies at least in part on the statistical regularities present around us, and often operates automatically and without any explicit, verbalizable knowledge of underlying rules \autocite{Aslin2012}. Statistical regularities can be inferred from various sources of information, including but not limited to the temporal order in which stimuli are experienced. As early as infancy, humans reliably detect the probabilities with which one stimulus transitions to another (\emph{transition probabilities}, such as one syllable following another in spoken language) and the frequencies with which stimuli temporally co-occur (\emph{co-occurrence frequencies}) \autocite{Saffran1996}. Similar forms of pattern sensitivity have been observed beyond the language domain, including motor learning \autocite{Nissen1987,Hunt2001} and visual event segmentation \autocite{Fiser2002,Turk-Browne2005}.

Ongoing research examines which types of statistics induce learning -- including statistical associations between movements \autocite{Cleeremans1991} and between non-linguistic sounds \autocite{Furl2011}. Moreover, evidence suggests that, depending on context, learners can extract both adjacent and non-adjacent dependencies between stimuli \autocite{Newport2004,Gomez2002}. Taken together, these studies suggest that second- or third-order statistical relationships may be encoded implicitly, and furthermore, that higher-level organizational principles themselves might be implicitly learned. Indeed, recent work has shown that temporal ordering of visual stimuli can convey the organizational principle of modularity \autocite{Schapiro2013}. This observation opens up the possibility of studying whether certain organizational principles are more or less facilitative of learning, and whether information embedded in certain organizational structures might be easier to learn than information embedded in others.

An ideally suited language in which to define such higher-order principles is network science \autocite{newman2010networks}, an emerging interdisciplinary field that addresses the architecture, dynamics, and design of complex systems composed of many connected parts \autocite{newman2011complex}. The set of parts (network nodes) and connections (network edges) are often parsimoniously encoded in a mathematical object called a graph \autocite{bollobas2001random}. In the context of learning, we can construct a graph that encodes the pattern of relationships between objects, movements, or sounds. Prior theoretical work \autocite{Jarvis1999} has addressed the relationship between graph structure and artificial grammars (such as in \authorcite{Cleeremans1991}), and we build on this work by empirically addressing the impact of graph-based properties. Recently \authorcite{karuza2017process} capitalized on this approach to define a graph from which the temporal ordering of visual stimuli was drawn. Learners exhibited a robust on-line measure of learned graph structure: a \emph{surprisal effect} defined as an increase in reaction time when transitioning between modules. Importantly, this surprisal effect was dependent on the type of traversal through the graph, and was more strongly pronounced when traversals through the graph provided redundancy in local information.

While the manner in which a graph is traversed can influence learning, the nature of the graph itself may serve as an even more fundamental constraint on the potential for humans to learn organizational principles of information. Many real-world systems including language \autocite{Goldstein2014}, conceptual knowledge \autocite{bales2006graph,vitevitch2008what}, social groups \autocite{palla2007quantifying}, and societies \autocite{Girvan2002} display non-trivial higher-order structure such as clustering or hub-and-spoke architecture that is relevant for how humans can optimally communicate, interact, and ultimately survive in their environment. Moreover, our knowledge about these systems unfolds and grows over time as we experience new parts (nodes) and their relations (edges). Understanding the impact of such higher-order structure on learning could help to explain why knowledge of some (natural or man-made) systems may be more easily acquired than others, and why individuals differ in their capacity to learn them. It may also shed light on the question of how humans can generalize from local statistical information to develop representations of broad-scale organizational patterns \autocite{karuza2016local}.

Here, we examined whether the higher-order regularities of three graph structures influenced implicit learning of statistical relationships among temporally ordered stimuli. Specifically, we trained subjects on a self-paced Serial Reaction Time (SRT) task, where each trial was drawn from a traversal through a graph. Each node represented a stimulus, and each edge represented a possible transition between two nodes. Based on prior work demonstrating learners' sensitivity to higher-order statistics in SRT-like tasks \autocite{Cleeremans1991}, we hypothesized that learners would display sensitivity to graph structure as evidenced by a surprisal effect. Next, we systematically varied graph structure to examine the impact of graph topology on the acquisition of complex, multi-element motor sequences. We hypothesized that learners would display increasingly rapid execution of button presses when presented with modular graph structures in comparison to either random or lattice graphs. The predicted preference for modular graphs is based on evidence across disparate fields of scientific inquiry. Modular topologies are more frequently observed in real-world systems than either random or lattice-like topologies \autocite{Steyvers2005}. Moreover, the clustering and hierarchy of modular graphs in natural systems can emerge in response to constraints on network wiring costs \autocite{mengistu2016evolutionary}, and similar constraints on complexity may impact learnability itself. Indeed, we predicted that learning mechanisms should be tuned to statistical features of natural stimuli \autocite{Hermundstad2014}. Finally, we hypothesized that learning performance would vary over different regions of the graph based on both local and global properties. Local properties are those inherent in a single node and its immediate neighbors, and global properties encompass organizational features of the entire graph such as clustering or repeated structure. Together these properties can be used to assess learners' sensitivity to variations in the graph across topological scales. Collectively, the results we report below suggest that learners extract and exploit the graph topology defining temporal sequences of stimuli, and that topological features impact speed of acquisition.

\begin{figure}[!ht]
\centering
\includegraphics{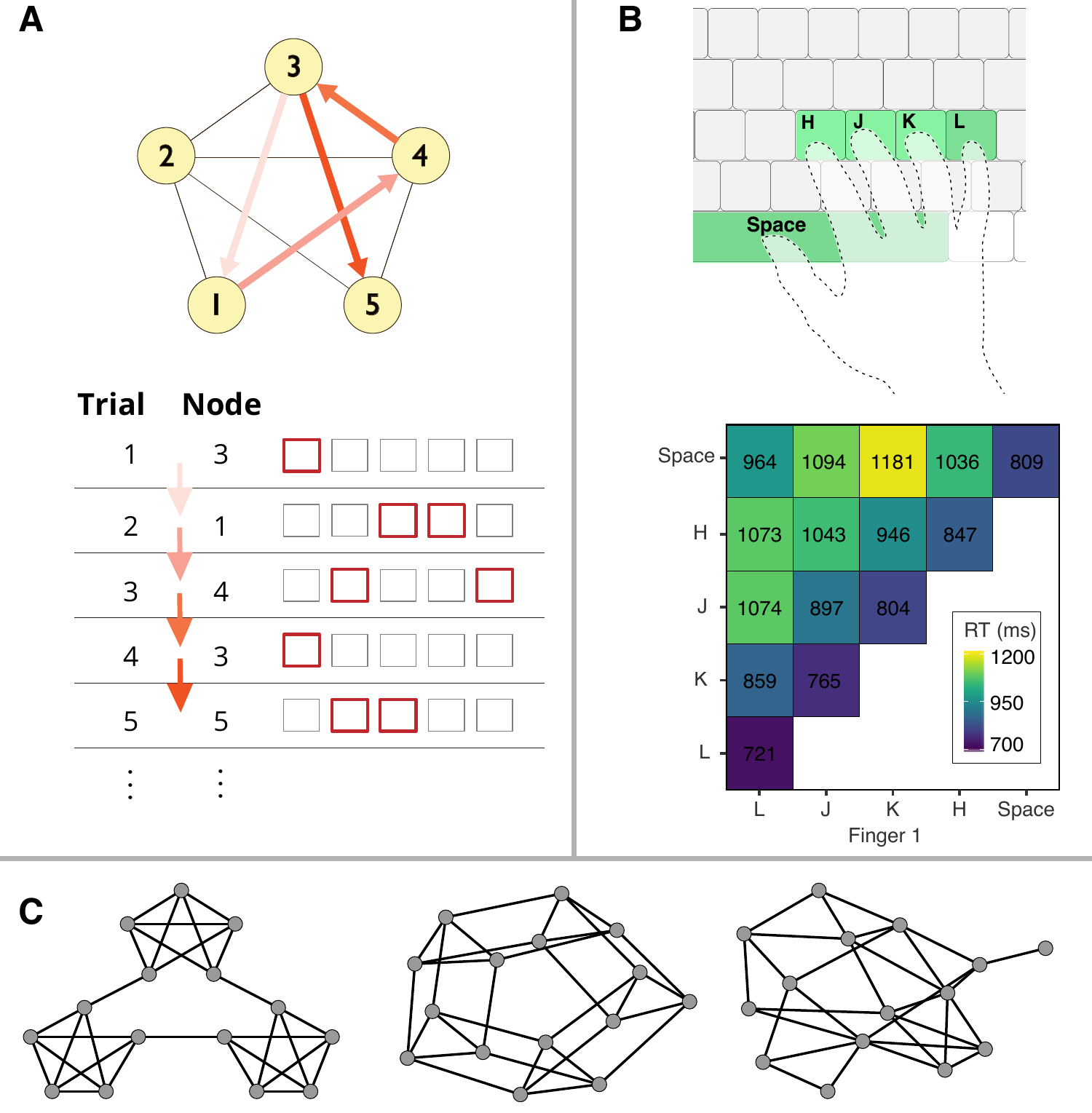}
\caption{\textbf{Experimental Setup.} \emph{(A)} An example of the first few steps of a graph traversal defined by a walk on the graph. \emph{Top:} Each node is uniquely associated with a key combination, and the sequence of key combinations is determined by a walk on the graph. \emph{Bottom:} A series of trials are presented to the participant. The red squares indicate which keys to press on that trial. Colored arrows illustrate the edge from the graph at  top being traversed. However, the participant only is shown the five squares. \emph{(B)} The mapping between fingers and keys, and average reaction times for each key press. \emph{Top:} A schematic of the mapping between visual stimuli (squares) and response effectors (fingers). \emph{Bottom:} The average reaction time (RT) for each key or pair of keys across all data. The diagonal elements of the matrix represent trials in which a single key was pressed, and the off-diagonal elements of the matrix represent trials in which a pair of keys was pressed. \emph{(C)} The three graph structures that we examine in this study. From left to right, we show a modular graph, a lattice graph, and a random graph with $N=15$ nodes connected by $E=30$ edges.
\label{figure1}}
\end{figure}

\section*{Results}

\textbf{Setup:} We recruited 381 unique participants: 109 participants for a first experiment, 59 participants for a second experiment, and 223 participants for a third experiment. Subjects performed a self-paced SRT motor response task using a keyboard. Stimuli were five grey squares in a horizontal row; to indicate a target key or pair of keys that the subject was meant to press, the corresponding square(s) would be outlined in red (Fig.~\ref{figure1}A). The squares corresponded spatially with keys `Space', `H', `J', `K' and `L', such that the left square represented `Space' and the right square represented `L' (Fig.~\ref{figure1}B). These keys were chosen so that the subject's hand could ergonomically rest over all five keys at once.
\textbf{Sequence Generation:} The order in which stimuli were presented to the subject was prescribed by either a random or a Hamiltonian walk on a graph of $N=15$ nodes connected by $E=30$ edges (Supplementary Table~\ref{experiment-table}). \textbf{Random Walk:} For any two nodes that \emph{were not} connected by an edge, the transition probability was equal to zero. For any two nodes that \emph{were} connected by an edge, the transition probability was equal to 1 divided by the number of edges emanating from the pre-transition node.
\textbf{Hamiltonian Walk:} A walk composed of a series of cycles, each of which visited every node on the graph exactly once. Each cycle was randomly generated starting from a node adjacent from the endpoint of the previous cycle. \textbf{Graphs:} We compared learning rates across 3 different graph topologies, each consisting of 15 nodes and 30 edges: a \emph{modular} graph, a \emph{lattice} graph, and a \emph{random} graph (Fig.1C). Briefly, the modular graph consists of three clusters of five interconnected nodes. The lattice can be thought of as a grid, wrapping around at its boundary. The random graphs differed between subjects and had no consistent organizational principles besides constituting a single connected component and maintaining the same number of nodes and edges as the other two graphs. (See Methods for formal definitions.) For both the modular and lattice graphs, the equal degree distribution coupled with a random walk leads to uniform pairwise probabilities for all possible transitions from a given node in the graph.
\textbf{Experiment:} We ran three experiments, each consisting of two back-to-back \emph{stages} that differed in which graph and walk type was used to generate the stimulus sequence.
The first experiment considered learning on one of three distinct topologies (\emph{modular}, \emph{lattice}, or \emph{random}) instantiated as sparse graphs containing only a minority of possible edges between nodes; learning from one of the structured topologies was followed by learning from a fully connected graph structure, in an effort to identify any changes in learning driven by the addition of novel edges. Pragmatically, the experiment was operationalized with a within-subjects design. Specifically, the first stage of the experiment used either a \emph{modular}, \emph{lattice}, or \emph{random} graph to generate a sequence of 1500 stimuli via a random walk on the graph. The second stage of the experiment used a fully connected graph to generate a sequence of 500 stimuli via a random walk on the graph (which amounted to a random stimulus order). With every node connected to every other node (eliminating any informative structure of the input), the fully connected graph allowed a comparison between previously trained edges and novel edges.
The second experiment consisted of 1500 trials of a random walk on a modular graph, followed by 500 trials of a Hamiltonian walk on the same graph, allowing us to confirm that learned differences transferred to a different walk type on the same graph structure.
The third experiment directly compared learning effects between graph types, accounting for individual variability in baseline reaction times and learning rates using a within-subjects design. Similar to the first experiment, the first stage of the experiment consisted of a sequence of 1500 stimuli via a random walk on either a modular graph, a lattice graph, or a random graph. However, unlike the first experiment, the second stage was another sequence of 1500 stimuli via a random walk on one of the remaining two sparse structured graph types. 
\textbf{Analysis:} We verified the surprisal effect using all modular graph traversals from stage 1 (Experiment 1). The effect of modifying the graph topology through the addition of novel edges was assessed using a sparse graph followed by a full graph in a between-subjects design (Experiment 1). Our assessment of the sensitivity of the surprisal effect to a switch from a random walk to a Hamiltonian walk was based on Experiment 2. Our comparison of behavioral performance between the sparse graph types (modular, lattice, and random) was based on a within-subjects design (Experiment 3). Finally, our comparison of local and global graph properties was based on data acquired during the random walk traversal of a sparse random graph (stage 1 of Experiments 1 and 3).

After determining differences in reaction time by key or key combination (Fig.~\ref{figure1}B) (which was then added as a regressor in all subsequent models; see Methods), we asked whether participants displayed evidence of learning the probabilistic motor sequence. A commonly studied marker of motor skill learning is an exponential drop-off in movement time with trials practiced \autocite{Heathcote:2000uy}. We observed this drop-off, despite our task being probabilistic rather than deterministic. Learners exhibited large decreases in reaction time on average over the course of the first experiment. Across the initial structured graph stage, we observed mean reaction time decreasing by nearly 500 ms (Fig.~\ref{figure2}A, black line. See Supplementary Figures~\ref{average_rt_by_experiment} \&~\ref{average_rt_by_graph} for additional information).

The observed overall decrease in reaction time with training suggested general improvement in task performance. To explicitly evaluate learning based on graph structure, we tested for the \emph{cross-cluster surprisal effect}, a previously reported measure of graph learning on modular graphs \autocite{karuza2017process}. Briefly, the cross-cluster surprisal effect is a significant increase in reaction time when a participant traverses an edge located between clusters, in comparison to when a participant traverses edges within clusters. Using a mixed effects model (see Methods), we observed a statistically significant increase in reaction time across edges that connected two clusters (Fig.~\ref{figure2}A,B; linear mixed effects model; $t(29)=3.61, p<0.002$, expected increase of 63.6 ms; 95\% confidence interval: 29.07 to 98.10, see Supplementary Table~\ref{transitional-rt-model}). To provide an intuitive visualization of this finding, we first note that the graph was symmetric across the three clusters; because the starting position and traversal direction within the graph varied for each subject, the distinction between the three clusters was arbitrary when comparing across subjects. We therefore remapped each edge to the equivalent edge within a single canonical cluster, which visually highlights the clear difference in reaction times for between- \emph{versus} within-cluster edges (Fig.~\ref{figure2}C).

\begin{figure}[!ht]
\centering
\includegraphics{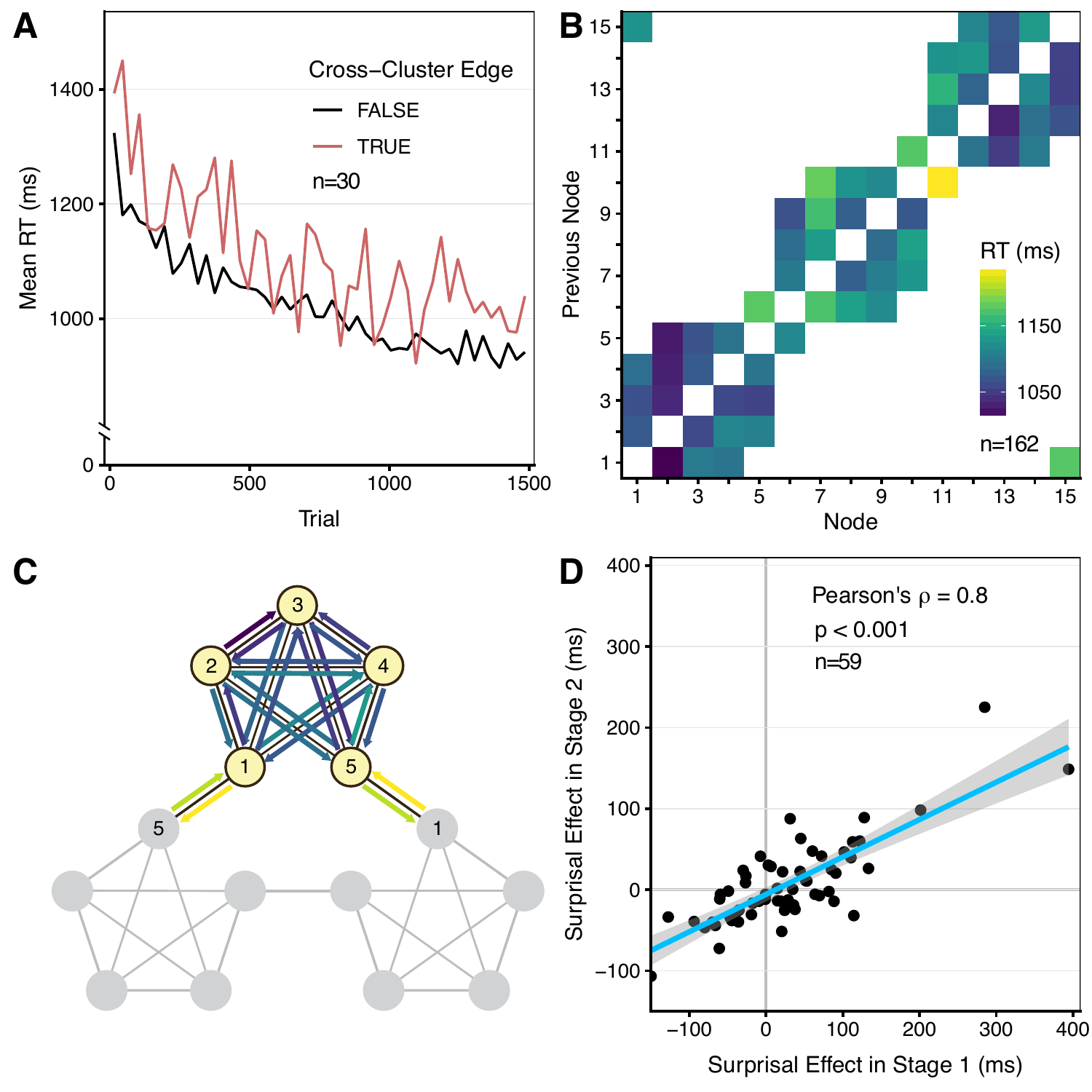}
\caption{\textbf{Modular Graph Learning Effects.} \emph{(A)} Mean reaction times (RTs) as a function of trial number for stage 1 of Experiment 1, among participants exposed to the modular graph. The red line indicates the mean for cross-cluster trials, and the black line indicates the mean for all other trials, each binned in sets of 30 trials (n=30 subjects). \emph{(B)} Mean reaction times on correct trials for the modular graph. An increase in reaction time across cluster boundaries can be seen, here visualized by yellower colors in the matrix elements that sit between the larger blocks. \emph{(C)} Mean reaction times collapsed across the symmetric structure of the modular graph. All three clusters were structurally identical and starting position was randomized between subjects, so we combine reaction times across the three clusters into one `canonical' cluster for visualization purposes only. The mean increase in reaction time between clusters is more apparent, here visualized by yellower colors on the edges that connect the top cluster with the two bottom clusters. \emph{(D)} Relationship between surprisal effect on stage 1 (random walk) and surprisal effect on stage 2 (Hamiltonian walk) for each subject of Experiment 2. Subjects that displayed a strong surprisal effect in stage 1 likewise do so when the walk structure is changed (n=59).
\label{figure2}}
\end{figure}

The observed surprisal effect suggests that participants are sensitive to graph structure. However, an alternative explanation is that the surprisal effect reflects a difference in processing cost inherent to local repetitions associated with a random walk on the modular graph. To either support or dismiss this alternative explanation, we asked whether the surprisal effect would persist when we modified the walk to sample sparsely from each module in time, eliminating within-module repetitions of button presses. Prior work \autocite{karuza2017process} suggests that learners are unlikely to show a surprisal effect when exposed to a Hamiltonian walk, where each node is only visited once per cycle. However, if clusters correspond to a learned feature of the graph, then we expect that learners first trained on a random walk (where we expect a surprisal effect) on a graph, and then exposed to a Hamiltonian walk (where we do not expect a surprisal effect) on the same graph, will continue to show the surprisal effect.

In this experiment, subjects were first exposed to 1500 trials of a random walk on the modular graph, followed by 500 trials of a Hamiltonian walk on the same graph. We were interested to determine whether a subject's surprisal effect in the first stage was correlated with their surprisal effect in the second stage. Using a mixed effects model, we found that our estimate for the surprisal effect was significantly reduced and not significant in the Hamiltonian walk, with an estimated increase in RT of 7.16 ms (Supplementary Table~\ref{surprisal-transfer-table}; linear mixed effects model; $t(58)=0.69, p=0.49$, 95\% confidence interval: -13.17 to 27.49). However, a subject's coefficient for cross-cluster surprisal in the random walk was significantly correlated with that in the Hamiltonian walk (Fig.~\ref{figure2}D; $t(57)=10.089, p<0.001$, Pearson's correlation coefficient $r=0.8$, 95\% confidence interval: 0.69 to 0.88), suggesting a diminished yet persistent effect of topological edge role after eliminating local repetitions. We verified that this result was not the result of idiosyncrasies in assignment of motor actions to nodes by performing a permutation test where the identity of cross-cluster edges was randomly assigned (see Supplementary Fig.~\ref{permute_random_hamiltonian}).

\begin{figure}
\centering
\includegraphics{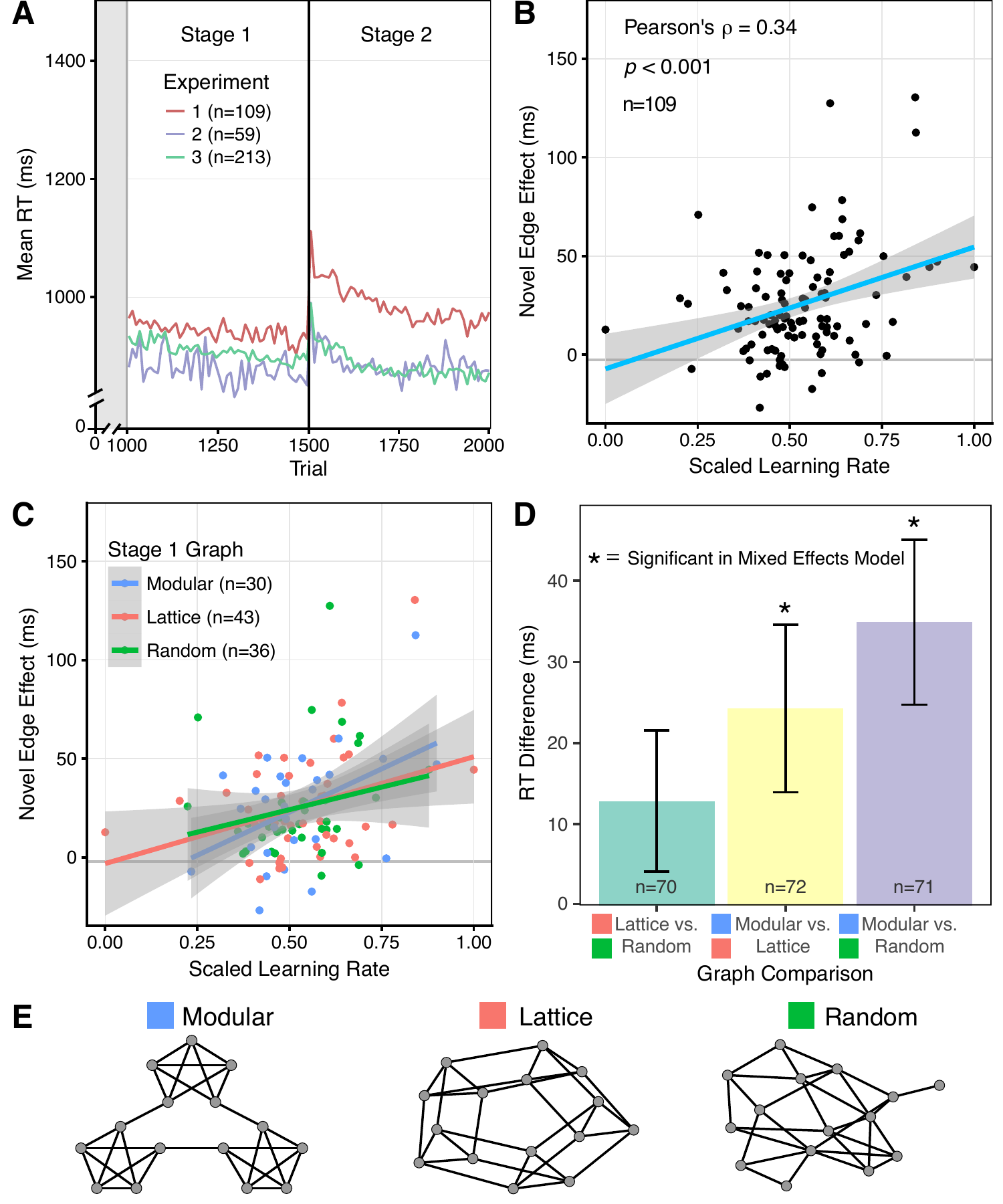}
\caption{\textbf{Learning Rate and Edge Surprisal.} Impact of new edges on reaction time. \emph{(A)} Mean RT increases in stage 2 when new edges are added to the graph (trials 1501-2000). Included for comparison are Experiments 2 and 3, where -- respectively -- only the walk or a subset of edges were changed. In both cases the increase in RT is much smaller. \emph{(B)} Per-subject learning rate correlated with the novel edge effect, defined as the mean difference in reaction time for a subject learning the second graph when responding to a novel edge \emph{versus} a familiar edge (see Methods; n=109). Learning rate, the model coefficient for log(trial), was scaled amongst all subjects to the range [0,1]. The blue line is the least squares fit, with the gray envelope indicating the 95\% confidence interval. \emph{(C)} Individual correlations shown for the three types of graphs trained on in the first stage. Subjects exposed to the modular and lattice graphs show a significant relationship ($p<0.01, n=30$ and $p<0.03, n=43$, respectively), while those exposed to the random graph do not ($p<0.2, n=36$). Solid lines represent least squares fits, and gray envelopes represent the respective 95\% confidence intervals. \emph{(D)} Differences in reaction time by graph type, across graphs learned in sequence. Each bar shows the number of milliseconds by which the modeled effect for the top listed graph is faster. The increase in RT from lattice to modular, and from random to modular graphs, are both significant to $p=0.02$ and $p=0.001$, respectively (See Supplementary Table~\ref{graph-effects-table}). Error bars indicate standard error as estimated in the mixed effects model. Asterisks indicate significance in the mixed effects model. Group sizes: Lattice-Random: n=70, Modular-Lattice: n=72, Modular-Random: n=71. \emph{(E)} Examples of the graph types.
\label{figure3}}
\end{figure}

While the cross-cluster surprisal effect is a useful measure of how well a modular graph is being learned, it is not a measure that generalizes to non-modular structures. To quantitatively examine the learnability of graph structure across many graph topologies, it would be useful to develop a generalizable measure of the learnability of single transitions from one button press to another. In developing such a measure, it is important to note that two potential explanations exist for improvement in response to a given target: (i) improvement is node-dependent (for all edges leading to that node), or (ii) improvement is edge-dependent, with the rate of improvement depending on the preceding node. Notably, the inclusion of additional edges (with the same set of nodes) in stage 2 of Experiment 1 led to a large increase in mean reaction time (Fig.~\ref{figure3}A). To verify that subjects had learned the edges \emph{versus} nodes of the graph, we examined stage 2 of Experiment 1: when subjects were shown (and asked to respond to) a sequence of stimuli drawn from a fully connected graph. We labeled each edge as either previously learned or not previously learned, based on the respective motor action assigned to the endpoint nodes and whether the edge (sequence of motor actions) was present in the graph learned by that subject in the first stage. We then estimated learner sensitivity to new edges as measured by a change in reaction time, which we referred to as the \emph{novel edge effect} (see Methods). Using a mixed effects model, we found that subjects were significantly slower when responding to edges that had not previously been seen, with an expected increase of 25.5 ms (linear mixed effects model; Supplementary Table~\ref{new-edge-surprisal}; $t(131)=4.35, p<0.001$, 95\% confidence interval: 14.01 to 37.0). This finding supports the notion that subjects learn single edges in a graph. However, altering the number of edges in the graph decreases predictability of transitions from a single node.

While subjects were faster on previously seen edges, this difference in reaction time could be attributable to improvements in compound motor movements, rather than to the learning of any higher-order structure. We therefore asked whether reaction time improvements for sequences generated by each graph type might modulate this novel edge effect. In other words, was it the case that subjects with greater sensitivity to graph structure would be more affected by disruptions to it? Further, might this association between learning measures differ by graph type? We estimated each subject's learning rate \autocite{Karuza2014}, and we also estimated the novel edge effect for each subject. We found that faster learners showed a significantly greater novel edge effect in the second stage of the experiment than slower learners (Fig.~\ref{figure3}B; $t(107) = 3.79, p<0.001$ Pearson's correlation coefficient $r=0.34$, 95\% confidence interval: 0.17 to 0.50). When subdivided by graph type (Fig.~\ref{figure3}C), this effect was significant for the modular graph ($t(28)=2.93, p=0.007$, Pearson's correlation coefficient $r=0.48$, 95\% confidence interval: 0.15 to 0.72) and the lattice graph ($t(41)=2.28, p=0.027$, Pearson's correlation coefficient $r=0.34$, 95\% confidence interval: 0.04 to 0.58), but not significant for the random graph ($t(34)=1.31, p=0.2$, Pearson's correlation coefficient $r=0.22$, 95\% confidence interval: -0.11 to 0.51). Intriguingly, this pattern of results suggests that subjects can more easily learn the regular structure of modular and lattice graphs, and display slower reaction times when expectations are violated. We note, however, that the difference \emph{between} the modular and lattice conditions and the random condition was not in itself significant (modular and random: Fisher's $z=1.18$, one-sided $p=0.12$, lattice and random: Fisher's $z=0.55$, one-sided $p=0.29$.)

Next, we tested whether certain graph structures facilitate learning more than others. We predicted that sequences generated by the modular graph would be the easiest for participants to learn, due to the graph's segregated meso-scale structure. As subject groups were exposed to different pairs of graph topologies, we performed three separate within-subject analyses using the data from Experiment 3. Each analysis examined a pair of graph types, with the order of exposure counterbalanced between subjects. For example, the first group was composed of (i) subjects first exposed to a stream of stimuli produced by a random walk on the lattice graph, followed by a stream of stimuli produced by a random walk on the random graph, as well as (ii) subjects first exposed to a stream of stimuli produced by a random walk on the random graph, followed by a stream of stimuli produced by a random walk on the lattice graph. In the same manner, the second group corresponded to modular/lattice, and the third group corresponded to modular/random. We separately fit a mixed effects model to each group. We found that the modular graph elicited significantly quicker responses than both the lattice (linear mixed effects model; $t(70)=2.35, p=0.022$; expected difference of 34.89 ms; 95\% confidence interval: -44.49 to -4.02) and random ($t(69)=3.429, p=0.001$; expected difference of -34.89 ms; 95\% confidence interval: -54.82 to -14.95) graphs (Fig.~\ref{figure3}D,E). We did not find a significant difference between the lattice and random graphs ($t(68)=1.48, p=0.14$; expected difference of 12.85 ms; 95\% confidence interval: -29.88 to 4.17). Models are summarized in Supplementary Table~\ref{graph-effects-table}. These findings support the hypothesis that the presence of meso-scale structure in modular graphs impacts learnability.

\begin{figure}
\centering
\includegraphics{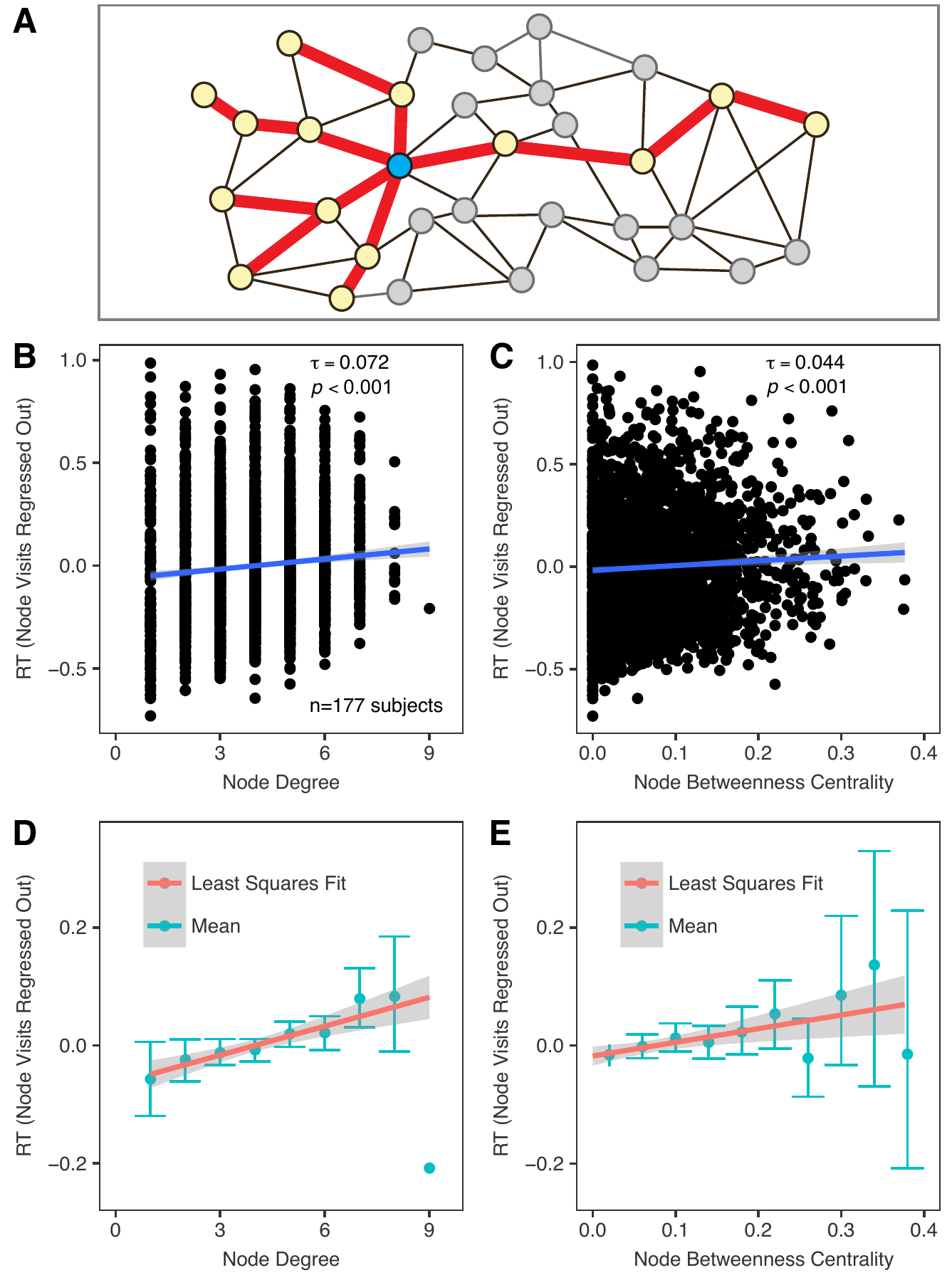}
\caption{\textbf{Relation Between Small and Large Scale Graph Statistics and Reaction Time.} \emph{(A)} Illustration of node betweenness centrality. We show shortest paths from a number of nodes on the far left to a node on the far right, which all pass through the blue node. \emph{(B)} Relationship between node degree and reaction time (RT), after regressing out the number of visits to a node, with each point representing a separate node, e.g., 15 points per subject. The regression line shows least squares fit, and the gray envelope is the 95\% confidence interval. Reported correlation is based on Kendall's $\tau$ (n=177 subjects). \emph{(C)} Relationship between node betweenness centrality and reaction time using the same approach as used with node degree. \emph{(D)} Mean reaction time shown as a function of degree, where the mean was $z$-scored across the 15 nodes for a given subject. Error bars represent bootstrapped 95\% confidence intervals. \emph{(E)} Mean reaction time as a function of node betweenness centrality using the same approach as used with node degree.}
\label{figure4}
\end{figure}

In a final set of analyses, we investigated the extent of the influence of graph structure on learning. More specifically, we tested whether smaller scale topological features or larger scale topological features might also impact learning, in addition to the meso-scale features studied in the previous section. First, we studied smaller scale topological features using degree, a summary statistic of a node's neighborhood defined by the number of edges emanating from a node. Second, we examined large scale topological features using betweenness centrality, which intuitively captures a node's role in mediating long distance traversals through the graph, and which is defined by the fraction of shortest paths that pass through a given node (Fig.~\ref{figure4}A). We studied the relation between reaction time and these statistics specifically in the random graph exposures from the first and third experiment, where we observed the greatest variability in degree and betweenness centrality over nodes. In all cases, we regressed out the visits to a node, to separate the influence of network topology from the influence of increased exposure. We found that node degree highly predicted the mean response time on a node (Fig.~\ref{figure4}B,D; Kendall's $\tau=0.072, n=2655, p<0.001$), as did node betweenness centrality (Fig.~\ref{figure4}C,E; Kendall's $\tau=0.044, n=2655, p<0.001$). These results indicate that not only does meso-scale graph organization affect learnability, but so do smaller scale topological features quantifying the number of edges in a node's immediate neighborhood, and larger scale topological features quantifying a node's role in long-distance traversals through the graph. Intriguingly, we observed an inverted relationship when we refrained from regressing out the number of visits to a node, a fact that highlights the complex relationship between topology and learnability. (For full results, and additional findings related to other graph metrics, see Supplementary Fig. 4-7. Also note that degree and node betweenness centrality were correlated in the graphs analyzed in this experiment -- Kendall's $\tau=0.669, n=2655, p < 0.001$ -- indicating that nodes with dense local connectivity also play an important role in long-distance traversals.)

In turning to a discussion of our results, we begin by grounding our experimental setup and findings in the context of prior literature. Then in later sections, we turn our attention to a discussion of more specific implications of graph architecture for learning. In this study, we capitalize on a rich history of using motor response times as a learning measure (e.g., \authorcite{Cleeremans1991}; for reviews see: \authorcite{cleeremans98_implic_learn}, \authorcite{robertson07_serial_react_time_task}). While those studies (and much of the artificial grammar learning work since \authorcite{reber67_implic_learn_artif_gramm}) have clear parallels to the present findings, our experiments diverge in the fundamental question they address. While we similarly generate sequences by ``walking'' along the edges of a graph, we additionally systematically manipulate the topological properties of the graph underlying motor responses. In contrast to the bulk of the artificial grammar learning literature, which largely relies on an arbitrary finite-state grammar, we instead apply organizational principles informed by graph theory to study biases in human learning.

In the current study, we employed tools from the field of network science to determine whether and how graph structure influences sequence learning. More specifically, our experiment built upon previous work demonstrating that modular graph organization influences visual statistical learning \autocite{Schapiro2013,karuza2017process}. We extended this line of inquiry to the motor domain by assigning unique button presses to nodes of a modular graph, generating sequences via a random walk on that graph, and asking whether learners exhibit similar behavioral effects at the boundaries between clusters of nodes. Indeed, we found that learners displayed a sharp increase in reaction time when transitioning from one cluster of button presses to another, indicating that they developed implicit expectations about the underlying topology of incoming sequential input. The sum of these results indicates that learners capitalize on modular structure in developing expertise in executing complex motor sequences. We further generalized our observations to other graph topologies, and showed that certain local-, meso-, and global-scale features of graph architecture are associated with higher learning rates than others, suggesting a critical impact of graph topology on motor skill acquisition.

One important and outstanding question is the following: What exactly forms the basis for the observed increase in RT when switching between clusters in the modular graph? One could imagine a scenario in which the importance of the topological role played by the cross-cluster edges would lead to an optimization of the associated motor movements, and a decrease (rather than an increase) in RT. Our expectation of an increase in RT was primarily based on prior observations in similar tasks, particularly the visual perception task reported in \authorcite{karuza2017process}. The fact that we observe the same increase in RT at cross-cluster edges in a motor task suggests that the surprisal effect is a general (rather than modality-specific) property of probabilistic sequential learning of modular structures. Yet, its presence across task modalities does not equate to a cause. Here we describe several possible mechanisms for the surprisal effect at a number of different explanatory levels.

One level of explanation begins with the underlying neural processes. Using a visual item rotation detection task based on a graph traversal, \authorcite{Schapiro2013} measured the distinguishability of clusters by asking subjects to mark natural `segmentation' points. By studying fMRI data acquired during the performance of the task, the authors found that patterns of BOLD activity were more similar for items within clusters than for items between clusters, despite the fact that visual stimuli were randomly assigned to nodes. Based on this work, it is natural to ask whether a similar mechanism might apply to the processing of targets for our motor response task. If so, then similar neural representations of nodes within a cluster might allow for quicker responses to targets that are nearby in representational similarity space. It would be interesting to test this possibility in a future study that combined neuroimaging with the behavioral assay we provide in our study.

Neural processes aside, we have referred to the increase in RT at cross-cluster edges as a surprisal effect based on prior work \autocite{karuza2017process}. Nevertheless, alternative interpretations -- beyond surprise -- also exist. For instance, the increase in RT when entering a new cluster might be construed as analogous to the sequence initiation cost commonly observed in SRT \autocite{Hunt2001} and DSP \autocite{Verwey2010} tasks, where increases in RT are observed at the beginning of separable chunks of responses. The idea that our observed surprisal effect may in fact represent a preparatory cost is certainly plausible. However, it remains an open question how a sequence initiation cost might generalize to a situation such as the current task in which subsequences are less structured. Indeed, it would be interesting in the future to assess evidence for a type of sequence initiation cost that reflects preparation for a regime of likely (rather than fixed) responses.

The surprisal effect also shows strong similarities to the so-called switch cost observed in the task switching literature \autocite{Kiesel2010}. In common parlance, the switch cost is an increase in response time when learners are prompted to switch tasks between trials, where a task might be reporting the color of a stimulus, or the magnitude of a number. Typically each task is conceived of as having its own stimulus-response mapping, which is similar to the discrete set of responses anticipated within each of our clusters. While many studies have demonstrated the persistence of the switch cost even when subjects anticipate task order, a few studies have importantly shown that a switch cost is still observed when participants are led to implicitly learn a task sequence \autocite{Koch2001,Gotler2003}, similar to the implicit presentation of modular structure in our current study. Moreover, the relationship between task and sequence has been shown to be hierarchical \autocite{Schneider2006}, where both task and sequence interact with one another. This observation provides a link to our current work where not only do the two interact, but the task (in the form of a cluster) is \textit{defined} by the sequence. In many real-world situations, tasks do not have explicitly defined subtasks nor clear boundaries between those subtasks, and natural divisions are only learned through experience. Thus, a comprehensive exploration of the flexible interplay between task and task sequence might be a promising direction for future research.

One final framing of interest relates to the dependence of RT on the local topology as measured by the degree and betweenness centrality of the random graph. Both of these metrics capture the distinctiveness of a node's role within the graph. The degree reflects the number of accessible nodes from a single node; the betweenness centrality reflects the likelihood that a node will be necessary to traverse when moving between any pair of nodes in the graph. Importantly, the modular graph also has important local structure, and the nodes connecting two clusters in the modular graph (boundary nodes) serve a distinct role from the nodes existing within a cluster. One possibility is that a similar mechanism underlies the modulation of RT by the cluster structure and the modulation of RT by the degree and betweenness centrality structure. For instance, boundary nodes in the modular graph exhibit much higher betweenness centrality (0.22) than within-cluster nodes (0.04). And indeed, in both the random graphs and the modular graphs, higher betweenness centrality is significantly associated with higher RTs when differences in degree are accounted for. Thus one could conceive of the surprisal effect as a direct result of these graph properties. However, our current experiment does not allow us to further explore this relationship, as subjects only learned a single modular graph with two distinct classes of nodes. A potential future direction would be to explore whether this is a general relationship across other graphs with modular structure.

The primary aim of this study was to examine differences in learners' sensitivity to distinct graph structures, while holding constant the process through which these graphs were traversed (i.e., via a random walk). By exposing participants to modular, lattice, and random graphs, we sought to ascertain how higher-level structure might aid or impede motor skill acquisition. Given that modularity is an essential organizational principle underlying such varied systems as music \autocite{gleiser2003community} and social networks \autocite{Girvan2002,tompson2018individual}, we anticipated a privileged role for this form of information structure, relative even to sequences generated from the highly structured lattice graphs. In an initial between-subjects experimental design, we provided evidence that learners tracked pairwise statistics, or edges linking nodes across all three graph structures. However, the extent to which learners displayed sensitivity to novel edges was predicted by learning rate only in the lattice and modular graphs. In other words, the overall timecourse of learning throughout exposure to sequences generated by a random graph was not associated with sensitivity to local transition statistics on subsequent measures. We therefore propose that graphs featuring regular structural organization (i.e., lattice and modular) might serve to boost knowledge of local regularities. 

Further, by capitalizing on a complementary within-subjects experimental design, we directly contrasted the acquisition of sequences generated by distinct graph structures. Compellingly, learning rates for the modular graph condition were significantly faster relative to both the lattice and random graph conditions. While the differences between the highly structured modular condition and the relatively unstructured random condition were perhaps to be expected, the differences between the modular condition and the lattice condition were uniquely insightful. In particular, nodes within the lattice and modular graphs were precisely equated in degree. They only differed in their meso-scale architecture, wherein neighboring nodes within the modular graph were densely interconnected with one another. While the lattice graph was also highly structured, it lacked this key organizational property, to the detriment of the learner, which demonstrates that the learned pairwise associations do not capture the full scope of learners' pattern sensitivity. Instead, we provide evidence that learners clearly benefit from modularity when it underpins the generation of complex motor sequences. Notably, our performance measure associated with modularity, the surprisal effect, persisted even when altering graph topology and the walk taken upon that topology. However, the effect observed when we considered an altered transition structure was significantly weaker than the effect observed when we considered the original transition structure. We believe that we have ruled out simple confounds, particularly in having shown that the relationship between reaction time on the random and Hamiltonian walks is specific to those edges that bridge clusters, and not an artifact of our analysis methods. However, ruling out these simple confounds is not wholly satisfying, and it remains an open question whether the weakening of the surprisal effect reflects limited training in the first stage of the experiment, or whether learners discard their previous response biases as they adapt to the new statistical structure of the second stage of the experiment. Regarding the first point, while sensitivity to statistical structure emerges in a short time frame, persistence and generalization to new contexts may require more extensive training. Likewise, it would interesting to retest our current experimental setup, but without a shift to the Hamiltonian walk, to ask whether the break itself disrupts previously learned statistics. We note that learned statistics are particularly sensitive to contextual shifts\autocite{Gebhart2009}, and therefore it is possible that the division between the two sections of the experiment was too explicit given the short timeframe. Fully addressing this possibility will require further data collection in future.

The graph-specific effects that we observed indicated that meso-scale graph architecture impacts learning. Yet, these data do not address whether meso-scale architecture alone is privileged, or whether both smaller and larger scale topological features also play a role in the learning process. Using the random graph exposures, which had the greatest variability in multiscale network architecture, we studied (i) a measure of small scale topological structure in the node degree, which captures the extent to which a given node is connected to other nodes in a network, and (ii) a measure of large scale topological structure in the node betweenness centrality, which incorporates information about the role of a node within the entire graph by measuring its importance in shortest paths between other node pairs. We found that nodes of higher degree and betweenness centrality were associated with lower reaction times, as might be expected simply from the fact that learners would be more frequently exposed to these nodes via a random walk through the graph. Unexpectedly, however, after regressing out the number of times each node was visited, these relationships were inverted such that nodes of higher degree and betweenness centrality were associated with higher reaction times. This finding has important implications for how we understand the impact of smaller and larger scale architecture on learning. Specifically, at small topological scales, we propose that when learners are exposed to sequences generated by a heterogeneous topology such as is present in random graphs, a trade-off exists between repeated exposure to a given node (i.e., due to high degree) leading to a lower reaction time, and the complex representation introduced by the high number of its neighbors leading to a higher reaction time. At large topological scales, we similarly propose that a trade-off exists between repeated exposure to a given node due to its location along shortest paths in the graph, and the complexity of the possible paths along which it could be visited.

In our current study, degree and betweenness centrality were significantly correlated with stimulus exposure. These nodes thus displayed a tradeoff between (i) familiarity, as modulated by the frequency of exposure, and (ii) uncertainty, as modulated by the probability of the future state being narrowly versus widely spread amongst motor actions. Understanding this potential tradeoff remains an important area for future work. One tractable strategy could be to consider other, non-random walks on the graph that would allow a node to exhibit both low familiarity and high uncertainty, or \emph{vice versa}. It could also be useful to consider graph topologies that would allow variation in betweenness centrality without impacting degree, so as to disambiguate the effect of one versus the other. For example, graphs could be constructed in which low-degree nodes serve as bridges between disconnected areas of the graph, thus having high betweenness centrality. Separately modulating different local statistics on the graph, as well as the type of walk used, could allow for an expanded understanding of the topological drivers of the observed RT variation.

The separability of perceptual and motor learning in this and similar tasks continues to be a matter of debate. For example, \authorcite{Deroost2006} found that, in the context of an SRT task, stimulus-stimulus (i.e., perceptual) learning was limited to simple deterministic sequences. After training participants on more complex probabilistic sequences, they did not find evidence of perceptual learning. The separability of perceptual and motor learning systems is a challenging issue and one that is not yet fully resolved (e.g., it is likely that one system bolsters the other). Unfortunately, because the present set of experiments was not designed to address this issue, we cannot make strong claims one way or the other. Thus, we elect to maintain the most conservative interpretation of the data possible: our task involves sequences of motor responses, so we frame our results as evidence of motor skill learning. However, teasing apart perceptual and motor learning under this framework is a fascinating area for future study. Especially when considering closely related prior work \autocite{Schapiro2013, karuza2017process}, we suggest that our observed pattern of results is likely reflective of domain-general processes.

We note a few methodological considerations that are particularly pertinent to this work. First, there exists a broad literature on the theory of graph structure as well as on structures found in natural stimuli. Here we sample only a small portion of possible graphs representing stimulus relationships. We study two stereotyped graphs, one with meso-scale clustering (modular) and one with no meso-scale clustering (lattice). However, graphs can exhibit other diverse topologies such as core-periphery structure, as well as other configurations of both high and low clustering beyond those tested. Moreover, the graphs we examine are all comprised of 15 nodes. While impractical for the current study, the relationships between natural stimuli might best be represented by graphs composed of hundreds or even thousands of nodes. Thus an open question is how these results generalize to both larger and more diverse networks. 
Second, we collected no personal information on Mechanical Turk participants. While we screened for eligibility using location and browser, we collected no information on handedness, age, or prior typing experience. However, as our regression models all incorporate per-subject baseline and learning rate effects, we expect minimal impact on our results from any between-subject differences. Third, there exist several limitations to using participants from Mechanical Turk, who might each be viewing the experiment on different browsers and with different levels of accuracy and speed in their internet connection. Fourth, it is possible that reaction time differences might be driven by recency priming. We know that processing times are reduced for a stimulus recently viewed by the learner, and that different nodes within a graph may be differentially affected. This is primarily a concern for modeling the cross-cluster surprisal effect, where transition nodes may have not been seen as recently as pre-transition nodes. However, \authorcite{karuza2017process} found that low level perceptual priming effects did not fully account for the observed cross-cluster surprisal effect. Moreover, all between-graph comparisons are dependent on reaction time across the entire graph rather than between different classes of nodes. We also note that excluding priming effects may be overly conservative, given that they may serve as a local cue to graph organization. Fifth and finally, our study does not address the question of how the exploration of graph structures in real stimuli is instantiated in the brain. For future work, it would interesting to consider recent evidence that co-occurrence of visual stimuli leads to increasingly similar neural representations in particular areas of human neocortex \autocite{Messinger2001,Li2008}, and that these same areas can encode associative distances between objects \autocite{Schapiro2013,Garvert2017}. These data suggest that an understanding of the sensitivity to the topological properties of graph structures may have important implications in future for an understanding of neural coding.

In conclusion, we note that our results highlight the importance of topological structure in learning from a complex environment. We first demonstrate that learning of higher-level statistics operates in the context of an SRT task. Related previous paradigms instead focused on perceptual learning tasks without this complex motor component. Thus, our results suggest that graph-based statistical learning mechanisms are unlikely to be modality-specific. Second, we have examined the impact of systematic differences in graph organization on learning. We find significant differences in subject performance on different graph types, despite identical numbers of stimuli and possible transitions in each graph. In particular, subjects show overall faster reaction times on sequences drawn from the modular graph. Lastly, we begin to explore why sequences drawn from these graphs may be easier or harder to learn, based on node-level statistics that topologically classify the node in relation to both its immediate neighbors and its global role within the graph. Understanding the degree to which these results generalize to other graph structures remains an important direction for future research, as well as understanding whether these results can provide insight on the organization of naturally occurring complex systems. Additionally, our current study does not address the impact of long-term learning on the surprisal effect. Whether the increase in RT persists or disappears as learners become better trained on the task may help distinguish between competing causes.

\section*{Methods}

\subsection*{Participants}

All participants provided informed consent as specified by the Institutional Review Board (IRB) of the University of Pennsylvania, and study methods and experimental protocols were approved by the IRB. We recruited 381 unique participants to complete our study on Amazon's Mechanical Turk, an online marketplace for crowdsourced work. Worker IDs were used to exclude any duplicate participants, both within and between the three experiments. The entire sample included 109 participants for the first experiment, 59 participants for the second experiment, and 213 participants for the third experiment. No statistical methods were used to pre-determine sample sizes but our sample sizes are similar to those reported in previous publications \autocite{karuza2017process,Schapiro2013}

All participants were financially remunerated for their time. In the first experiment, participants were paid up to \$7 for an estimated 40 minutes: \$2 for completing each of the two stages, \$1 for completing the entire task, and an extra \$1 on each stage on which they correctly answered at least 90\% of trials. Experiment 2 provided the same payment as Experiment 1, but with an estimated duration of 40 minutes. In the third experiment, subjects were paid up to \$11 for an estimated 60 minutes: \$3 per stage, \$1 for completing the entire task, and \$2 for >90\% performance on each stage.

\subsection*{Experimental Setup}

Subjects performed a self-paced SRT motor response task using a keyboard. Stimuli were represented as a horizontal row of five gray squares; all five squares were shown at all times during the main phase of the experiment. To indicate a target key or pair of keys that the subject was meant to press, the corresponding squares would be outlined in red (Fig.~\ref{figure1}A). When subjects pressed the correct key combination, the squares on the screen would immediately display the next target. If an incorrect key was pressed, or a key was left out of a two-key combination, the message ``Error!'' was displayed on the screen below the stimuli, and remained until the subject pressed the correct key(s). The squares corresponded spatially with keys `Space', `H', `J', `K' and `L', such that the left square represented `Space' and the right square represented `L'. These keys were chosen to ergonomically rest underneath the subject's right hand on a QWERTY keyboard with their thumb above `Space', index finger above `H', and so on (Fig.~\ref{figure1}B).

The order in which stimuli were presented to the subject in Experiments 1 and 3 was prescribed by a random walk on a graph of $N=15$ nodes connected by $E=30$ edges. In each graph, one of the 15 one- or two-finger key combinations was randomly assigned to each node (Fig.~\ref{figure1}A). A different graph (mapping of key presses to nodes) was generated for each random walk. For any two nodes that \emph{were not} connected by an edge, the transition probability was equal to zero. For any two nodes that \emph{were} connected by an edge, the transition probability was equal to 1 divided by the number of edges emanating from the pre-transition node. In Experiment 2, stage 1 consisted of a random walk as previously described. The order of stimulus presentation in stage 2 was generated by a series of Hamiltonian cycles through the graph. A single cycle consisted of every node in the graph being visited exactly once, and each cycle was followed by another Hamiltonian cycle beginning from a node adjacent from where the last cycle ended.

We studied the learning of 3 different graph topologies: a \emph{modular} graph, a \emph{lattice} graph, and a \emph{random} graph (Fig.~\ref{figure1}C). The modular graph was characterized by 3 clusters of 5-nodes each, and a greater number of edges between nodes in a cluster than between nodes in different clusters. Importantly, each node in the graph had exactly 4 edges, or a degree of $k=4$. Thus, for any two nodes that \emph{were} connected by an edge, the transition probability was equal to 25\%. The lattice graph was similar to a ring lattice, in which nodes near one another on the ring tended to be connected to one another. As with the modular graph, each node had exactly 4 edges, or a degree $k=4$, thereby creating a flat transition probability of 25\% between any two connected nodes. Random graphs were selected out of a possible pool formed by creating 1500 instantiations of the Erdős–Rényi graph model, all of which we guaranteed were fully connected and had radius of at least 3, meaning there was at least one pair of nodes with a shortest path involving three edges. We then sorted the ensemble by their estimated modularity \autocite{newman2006modularity}, and we discarded graphs with the highest and lowest 2.5\% of modularity values. The same pool of random graphs was used across all subjects within a given experiment, though the pool differed between the experiments. In these random graphs, the degree of each node varied from $k=1$ to $k=9$, and the transition probabilities varied accordingly. Finally, in the second stage of the first experiment, we used a \emph{fully connected} graph as a point of comparison, in which any node can transition to any other node. In this case, the degree of each node was $k=14$, and the probability of transitioning between any two nodes was 1/14 or approximately 7.14\%.

We ran three separate experiments (see Supplementary Table~\ref{experiment-table}). Each one consisted of two \emph{stages} that differed in which graph was used to generate the stimulus sequence. The first experiment examined learning as the underlying structure transitioned from a regular graph to a fully connected graph, and did so by comparing reaction times in response to novel \emph{versus} previously learned edges. The first stage of the first experiment used either a modular graph (n=30), a lattice graph (n=43), or a random graph (n=36) to generate a 1500-node random stimulus walk, while the second stage used a fully connected graph to generate a 500-node walk (random stimulus order). In the second experiment, both stages consisted of a walk over the modular graph. However, the first stage was a 1500-node random walk, while the second stage was a 500-node Hamiltonian walk (n=59). The third experiment employed a within-subjects design to directly compare learning effects between the graph types, accounting for individual variability in baseline reaction times and learning rates. Similar to the first experiment, the first stage of the second experiment consisted of a 1500-node random stimulus walk from either a modular graph, a lattice graph, or a random graph; however, unlike the first experiment, the second stage was a 1500-node random stimulus walk on one of the remaining two graph types. For example, if a subject was shown a sequence of stimuli produced from the modular graph in the first stage, then in the second stage the subject would be shown a sequence of stimuli produced from either the lattice graph or the random graph. For each of the six possible pairs of graphs we collected reaction time data from at least 30 subjects (modular/lattice: n=36, modular/random: n=37, lattice/modular: n=36, lattice/random: n=38, random/modular: n=34, and random/lattice: n=32). In all cases, subjects were randomly assigned to experimental conditions. Assignment was done in the experiment code, blinding experimenters to the condition assignment for each individual participant. Subjects were only excluded if they failed to complete the study.

\subsection*{Experimental Procedures}

At the beginning of the first experiment, subjects were provided with the following instructions: ``In a few minutes, you will see five squares shown on the screen, which will light up as the experiment progresses. These squares correspond with keys on your keyboard, and your job is to watch the squares and press the corresponding key when that square lights up. This part will take around 30 minutes, followed by a similar task which will take about 5 minutes.'' To incentivize accuracy on the task, subjects were informed that if they answered more than 90\% of trials correctly, they would receive a \$2 bonus. Subjects were also instructed ``The amount of time the segments take is not fixed, but the number of responses you have to make is. Therefore, you should make your responses both quickly and accurately.'' While the reward was solely based on accuracy, workers had an implicit incentive to finish quickly due to the fixed reward, allowing more time for other tasks on Mechanical Turk.

Before the full experiment began, subjects were given a short quiz to verify that they had read and understood the instructions. If any questions were answered incorrectly, subjects were shown the instructions again and asked to repeat the quiz until they answered all questions correctly. Next, all subjects were shown a 10-trial segment that did not count towards their performance; this segment also displayed text on the screen explicitly telling the subject what keys to press on their keyboard. Subjects then began the 1500-trial stage. A brief reminder was presented before the second stage, but no new instructions were given. After completing the second stage, subjects were presented with performance information and their bonus earned, as well as the option to provide feedback. The second experiment used the same instructions as the first experiment, though the estimated time for the second task was changed to 10 minutes.

The third experiment was nearly identical, except that the initial text was changed to reflect the second structured walk: ``In a few minutes, you will see five squares shown on the screen, which will light up as the experiment progresses. These squares correspond with keys on your keyboard, and your job is to watch the squares and press the corresponding key when that square lights up. This part will take around 30 minutes, followed by a similar task which will take another 30 minutes.'' This phrasing in the instructions ensured that learners would differentiate between stages of the experiment, reducing the potential of carry-over effects between learning of graph structures while still preserving the benefit of a within-subject comparison. Both the quiz and 10-trial practice session were still present, and stimulus presentation was identical to that used in the first experiment.

Mixed effects models were fit using the lme4 library in R (R v3.5.0, lme4 v1.1-17), using the lmer function. Predictors were centered to reduce multicollinearity. All contrasts were orthogonally coded. The observed correlation between fixed effects was less than 0.7. Random effects were chosen as the maximal structure that allowed model convergence, as specified in the next section. All tests were two-sided unless otherwise noted.

\subsection*{Analytical Approach}

For every trial, we computed the reaction time based on the elapsed time from the last button press. We only excluded trials for two reasons: subjects answered incorrectly on their first attempt or the reaction time was implausible (under 100ms or over 5000ms, or more than 3 SDs from their mean reaction time).

\textbf{Effect of Targets on Reaction Time:} Since performance was measured based on a key press, it was important to determine whether biomechanical factors related to the use of different fingers, or to different combinations of finger pairs, influenced reaction time. We predicted that one-finger responses would show shorter reaction times than two-finger responses, and that response time would vary based on the finger needed in the response. We calculated the average reaction time for each key press or combination of key presses across subjects and training stages (Fig.~\ref{figure1}C). While we observed complex differences in reaction time by finger, we also found robust differences in reaction time driven by the number of fingers required, with one-key presses displaying a shorter reaction time than two-key presses (paired two-tailed $t$-test for one- and two-finger means for each subject: $t(321) = 35.56, p<0.001$; mean difference: 228.86 ms, 95\% confidence interval: 216.20 to 241.52 ms). We observed that the relative ordering of keys or key combinations by reaction time was remarkably well-preserved across subjects, and the difference was independent of which graph was seen in the first stage of the experiment (random, lattice, or modular). This observation suggests unique motor timing associated with each target (Supplementary Fig.~\ref{fingers_across_graphs}). To ensure that our findings were not systematically biased by differences in reaction time across key presses, we included the target key press as a regressor in all statistical models.

\textbf{Learning Rate:} Based on a subject's reaction time profile across the session, we estimated each subject's learning rate from a linear mixed effects model. The learning rate estimate was the per-subject random effect for trial number. A faster learning rate was captured as a negative coefficient, which indicated that a subject's reaction time decreased more rapidly over time. The model was fit to the first stage of the first experiment, with formula $\textit{RT} \sim \log(\textit{Trial})*\textit{Graph} + \textit{Target} + (1 + \log(\textit{Trial})\ |\ \textit{Subject})$, where \textbf{Target} represented the key combination of the target node, controlling for biomechanical differences in motor response, \textbf{Graph} was one of random/lattice/modular, and \textbf{Trial} was the sequential trial number (1 to 1500). We verified experimentally that $\log(\textit{Trial})$ provided a substantially better fit than $\textit{Trial}$ (See Supplementary Fig.~\ref{trial_log_trial}). The log transformation served to increase the normality of the data, although formal testing of the degree of normality was not performed. 

\textbf{Surprisal Effect:} One measure reflective of meso-scale graph structure is the \emph{surprisal effect}, defined as an increase in reaction time when transitioning to a new cluster as compared to any previous within-cluster reaction times \autocite{karuza2017process}. To measure this effect in our task, we fit a linear mixed effects model of the form $\textit{RT} \sim \log(\textit{Trial})*\textit{EdgeType} + \textit{Target} + (1 + \log(\textit{Trial})*\textit{EdgeType} \ |\ \text{Subject})$ to data acquired during the first stage of the experiment on the modular graph where $RT$ is reaction time, and where \textbf{EdgeType} indicated whether an edge was within or between clusters. From this model, we examined the model coefficient for \textbf{EdgeType}.

\textbf{Surprisal Transfer:} We fit the surprisal effect model separately to the stage 1 (random walk) and stage 2 (Hamiltonian walk) data from Experiment 2. We estimated the per-subject effect of \textbf{EdgeType} as a measure of surprisal, and additionally examined the Pearson correlation coefficient between each subject's surprisal effect in stage 1 \emph{versus} surprisal effect in stage 2.

\textbf{Novel Edge Effect:} Next, we sought to examine whether subjects were specifically improving their performance at transitions present in the graph and therefore sensitive to violations of the learned structure of the graph. We defined a \textbf{LearnedEdge} variable for all edges in stage 2 that was true if an edge between the set of finger combinations had been present in the first stage, and false otherwise. We then computed a \emph{novel edge effect} measure as the coefficient for learned \emph{versus} unlearned edges in a linear mixed effects model fit to the fully connected graph data using $\textit{RT} \sim \log(\textit{Trial})*\textit{Graph}*\textit{LearnedEdge} + \textit{Target} + (1 + \log(\textit{Trial})*\textit{LearnedEdge}\ |\ \textit{Subject})$.

\textbf{Graph Effects:} In the third experiment, we investigated whether differences in meso-scale structure affected learnability by quantifying reaction time differences due to graph type. Since not all subjects were exposed to all graph types, we split the data into three groups based on exposure: modular/lattice, random/modular, and random/lattice, with each group roughly split by which graph was used for the first stage \emph{versus} for the second stage. We fit a linear mixed effects model to each of the three data subsets, in order to estimate whether graph type was a significant effect in each model: $\textit{RT} \sim \log(\textit{Trial})*\textit{Graph}*\textit{Stage} + \textit{Target} + (1 + \log(\textit{Trial})*\textit{Graph} | \textit{Subject})$, where \textbf{Trial} varied between 1 and 1500, \textbf{Graph} was the current graph of the two graphs that the subject saw, and \textbf{Stage} corresponded to either the first stage or the second stage.

\textbf{Node Effects:} After determining the effect of graph type, we finally turned to quantifying the impact of node-level statistics, or those that could explain reaction time differences within a single graph. We examined several traditional graph metrics: degree, clustering coefficient, node betweenness centrality, and edge betweenness centrality. The degree is defined as the number of edges connecting to a given node, given by $k_i=\sum_j A_{ij}$ where $A$ is the adjacency matrix. The clustering coefficient can be defined as the fraction of possible edges between a node's neighbors, given by $C_i=\frac{2L_i}{k_i(k_i-1)}$ where $L_i$ is the number of edges between any two neighbors of node \emph{i}. The node betweenness centrality is defined as the fraction of shortest paths in the graph that pass through node $v$, given by $C_B(v)=\sum_{s\neq v \neq t}\frac{\sigma_{st}(v)}{\sigma_{st}}$ where ${\sigma_{st}(v)}$ is the number of shortest paths from node \emph{s} to node \emph{t} that pass through node \emph{v}, and ${\sigma_{st}}$ is the total number of shortest paths in the graph from node \emph{s} to node \emph{t}. The edge betweenness centrality is defined as the fraction of shortest paths passing through an edge \emph{e}, given by $C_B(e)=\sum_{s\neq e\neq t}\frac{\sigma_{st}(e)}{\sigma_{st}}$, where now \emph{e} refers to an edge rather than a node.

\section*{Data Availability}
The data that support the findings of this study are available from the corresponding author upon reasonable request.

\section*{Code Availability}
The code that supports the findings of this study is available from the corresponding author upon reasonable request.

\newpage

\printbibliography

\FloatBarrier

\section*{Acknowledgments}
We thank David M. Lydon-Staley and Steven H. Tompson for feedback on earlier versions of this manuscript. This work was supported by the National Science Foundation CAREER award to DSB (PHY-1554488), the Army Research Laboratory through contract number W911NF-10-2-0022, and the Army Research Office through contract number Grafton-W911NF-16-1-0474 and contract number DCIST- W911NF-17-2-0181. We also acknowledge additional support from the John D. and Catherine T. MacArthur Foundation, the Alfred P. Sloan Foundation, the ISI Foundation, the Paul Allen Foundation, the Army Research Office (Bassett-W911NF-14-1-0679), the Office of Naval Research, the National Institute of Mental Health (2-R01-DC-009209-11, R01 MH112847, R01-MH107235, R21 MH-106799, R01 MH113550), the National Institute of Child Health and Human Development (1R01HD086888-01), the National Institute of Neurological Disorders and Stroke (R01 NS099348), and the National Science Foundation (BCS-1441502, BCS-1631550, and CNS-1626008). The content is solely the responsibility of the authors and does not necessarily represent the official views of any of the funding agencies. The funders had no role in study design, data collection and analysis, decision to publish or preparation of the manuscript.

\section*{Competing Interests}
The authors declare no competing interests.

\section*{Author Contributions}
A.E.K., E.A.K., J.M.V., and D.S.B. conceived the project. A.E.K., E.A.K., J.M.V., and D.S.B. planned the experiments and analyses. A.E.K. performed the experiments and analyses. A.E.K., E.A.K., and D.S.B. wrote the manuscript and Supplementary Information. E.A.K., J.M.V., and D.S.B. edited the manuscript and Supplementary Information.

\end{document}